\documentclass[english,10pt,a4paper,nofootinbib,twocolumn]{revtex4}

\usepackage{amsmath,amssymb,amsfonts,amsthm}
\usepackage{epsfig}
\usepackage{graphicx}
\usepackage{latexsym}
\usepackage[colorlinks=true,linkcolor=blue,citecolor=blue]{hyperref}
\usepackage{subfig}
\usepackage{caption}    

\usepackage{tikz}
\usepackage{braids}
\usetikzlibrary{backgrounds,fit,decorations.pathreplacing}  
\usetikzlibrary{circuits.logic.US}

\makeatletter

\@ifundefined{textcolor}{}
{%
\definecolor{BLACK}{gray}{0}
\definecolor{WHITE}{gray}{1}
\definecolor{RED}{rgb}{1,0,0}
\definecolor{GREEN}{rgb}{0,1,0}
\definecolor{BLUE}{rgb}{0,0,1}
\definecolor{CYAN}{cmyk}{1,0,0,0}
\definecolor{MAGENTA}{cmyk}{0,1,0,0}
\definecolor{YELLOW}{cmyk}{0,0,1,0}
\definecolor{PURPLE}{rgb}{0.5,0,0.5}
}

\newcommand{\bite}{\begin{itemize}}
\newcommand{\eat}{\end{itemize}}
\newcommand{\beq}{\begin{equation}}
\newcommand{\eeq}{\end{equation}}

\newcommand{\beqa}{\begin{eqnarray}}
\newcommand{\eeqa}{\end{eqnarray}}
\newcommand{\barr}{\begin{array}}
\newcommand{\earr}{\end{array}}

\newcommand{\mb}[1]{\mathbf{#1}}
\newcommand{\mc}[1]{\mathcal{#1}}
\newcommand{\mbb}[1]{\mathbb{#1}}
\newcommand{\mf}[1]{\mathfrak{#1}}

\newcommand{\onehalf}{\frac{1}{2}}

\newcommand{\expect}[1]{\langle #1\rangle}

\newcommand{\fullket}[1]{\ensuremath{\left|#1\right\rangle}} 
\newcommand{\supersc}[1]{$^{\textrm{#1}}$}

\newcommand{\sltwoc}{\mathfrak{sl}(2,\mathbb{C})}

\makeatother

\begin{document}


\title{Elementary Particles as Gates for Universal Quantum Computation}

\date{\today}

\author{Deepak Vaid}
\email{dvaid79@gmail.com}
\affiliation{}

\begin{abstract}
It is shown that there exists a mapping between the fermions of the Standard Model (SM) represented as braids in the Bilson-Thompson model, and a set of gates which can perform Universal Quantum Computation (UQC). This leads us to conjecture that the ``Computational Universe Hypothesis'' (CUH) can be given a concrete implementation in a new physical framework where elementary particles and the gauge bosons (which intermediate interactions between fermions) are interpreted as the components of a quantum computational network, with the particles serving as quantum computational gates and the gauge fields as the information carrying entities.
\end{abstract}

\maketitle
	
\tableofcontents


\section{Introduction}

There is a certain prejudice among physicists which favors a bottom-up approach to understanding physical processes. Of course, all our understanding comes about from a physical experimentation process which generally proceeds top-down - i.e. the gradual disassembly of a system in order to understand how it might have formed in the first place. This is the process that led us to our present faith in the robustness of the description of elementary particles as irreducible representations of the continuous Lie groups corresponding to the observed symmetries of space-time i.e. the Poincare and Lorentz groups. However, the limits of this approach have become clear over time. Despite valiant efforts (such as grand unification via SO(5), SO(10), SUGRA, String/Matrix models etc.) physicists have been unable to find a clean, unified description of elementary particles which can reproduce not only the verified Standard Model Lagrangian $L_{SM}$ at low-energies $E \ll E_{planck}$ but also provide explanations for various phenomenological puzzles such as \emph{neutrino masses and mixing amplitudes, the number of generations of fermions, the hierarchy problem} among others. The present state of affairs in elementary particle physics is known as the ``Standard Model'', with the accompanying charges, masses and mixing angles which are \emph{free} parameters determined only by comparison with experimental results. 

We are therefore compelled to ask if our present understanding of physical law is fundamentally incomplete and requires an extension above and beyond any given prescriptions for a theory of quantum gravity such as loop gravity or `M'-theory.

Starting with Konrad Zuse's efforts in the earlier half of the twentieth century until the present day a consensus has been emerging about the nature of physical reality in which a computational paradigm plays a fundamental, and not merely supporting, role. Recent works by Seth Lloyd \cite{Lloyd1999Universe,Lloyd2006Theory}, Jurgen Schmidhuber \cite{Schmidhuber1999Computer} and Stephen Wolfram \cite{Wolfram1983Statistical}, among many others \cite{DAriano2010Physics,DAriano2010The-Quantum,Frieden1989Fisher,Frieden1990Fisher,Frieden1995Erratum:,Frieden1995Lagrangians,Frieden2002Black}, have added to the chorus of voices which appear to be arguing for a picture of physical reality not unlike that portrayed in the motion picture \emph{The Matrix}, though substantially less grounded in pop culture and more so in fundamental considerations about our notions of complexity and the equivalence of computational methods and physical law.

From a more pragmatic point-of-view, such considerations are suggested by the connection between topological quantum computation - particularly as implemented in a two dimensional electron gas (2DEG) via non-abelion anyons \cite{Nayak2008NonAbelian}, the problem of Black Hole entropy and the Holographic principle. Any such 
scheme would, however, be considered incomplete unless it presented us with a clear understanding of the role that particles of the Standard Model and the gauge interactions between them play in a computational setting.

If a strong connection between the physics of elementary particles and the processes of quantum computation can be established, then it would become less controversial to make arguments in support of the computational universe paradigm. Indeed, if such a mapping exists, it would naturally lead us to a picture of physical reality where the topological structure of space-time (i.e. constructed of braids, loops, strings etc.) can be thought of as hardware, the software corresponding to which would be topological rules governing basic physical processes such as the scattering of elementary particles.

In particular one would like to have a clearer understanding of how the different constituents of matter can play the different roles of information carriers and computational gates which govern how this information is processed. We argue that the discovery by Bilson-Thompson \cite{Bilson-Thompson2005A-topological} of a mapping between elements of the \emph{framed} braid group on three-strands $\mf{B}_3$ and the first generation of fermions of the SM, in conjunction with the work of Kauffmann, Lomonaco and Samuel \cite{Kauffman2004Braiding} demonstrating that the elements $B_3$ (the three-strand braid group, without framing) provide us with the gates needed for universal quantum computation, allows us to answer this question in a concrete manner.

The outline of this paper is as follows. Section \ref{sec:information} discusses the many parallels between information theory and physical science that have been uncovered over the past century. Section \ref{sec:gates} summarizes the minimal requirements for universal quantum computation. In Section \ref{sec:topo-comp} briefly covers the braid model of quantum computation. The Yang-Baxter equation is described in Section \ref{sec:yang-baxter} . In Section \ref{sec:bilson} we review the Bilson-Thompson model of topological preons and in Section \ref{sec:processors} we describe how the synthesis of these the preon model with the braid model of quantum computation provides us with an interpretation of particles and fields as processors and channels of a computational machine. We conclude with a discussion of the implications of this new perspective for our understanding of physical processes involving elementary particles. An appendix \ref{sec:large_gauge} describing the emergence of braiding in spin-networks in the framework of loop quantum gravity is included.


\section{Information and Theoretical Physics}\label{sec:information}

Beginning with the pioneering works of Jaynes, Landauer, Bennett, Bekenstein, Hawking, Unruh, Davies and leading upto more recent breakthroughs by Jacobson, Maldacena, Bousso and others, it has become clear that information plays more than just a supporting role in the foundations of physics. Jaynes \cite{Jaynes1957Information} was the among the first to note that statistical mechanics could be constructed from information-theoretic considerations. Landauer \cite{Landauer1961Irreversibility,Landauer1996The-physical}, first established the intimate relationship between computation and thermodynamics by arguing the erasure of a single bit of information from a computational system results in an irreversible increase in the entropy of the system by an amount $ kT $, where $ k $ is Boltzmann's constant and $ T $ is the characteristic temperature of the system.

Later, Bekenstein \cite{Bekenstein1973Black} showed that the irreversible nature of certain black hole transformations - in the case of a rotating (Kerr) black hole as first investigated by Christodoulu - implied that a black hole could be thought of as a thermodynamic system whose entropy was proportional to the area of the black hole horizon:
\begin{equation}\label{eqn:bekenstein-bound}
	S_{BH} = \alpha A
\end{equation}
where $ \alpha $ is a universal constant. Information theory played a crucial role in Bekenstein's original considerations \cite{Bekenstein2003Black} though that aspect appears to have been de-emphasized in later work on black hole entropy. The developments associated with the holographic principle and Maldacena's AdS/CFT correspondence in the late 1990s brought the centrality of information in theoretical physics back into focus.


The physical significance of the so-called Bekenstein bound \ref{eqn:bekenstein-bound}, lies in the fact that it places an upper limit on the entropic content and therefore on the maximum amount of information that can be stored in \emph{any} finite region of spacetime.

In 1995, Ted Jacobson showed \cite{Jacobson1995Thermodynamics} that one could \emph{derive} the Einstein field equation assuming only that the second law holds near all Rindler horizons. The thermal origin of gravity was extended and improved upon by Padmanabhan \cite{Padmanabhan2002Classical,Padmanabhan2002The-Holography,Padmanabhan2003Gravity,Padmanabhan2009A-Physical} through the first decade of the $21$\supersc{st} century. the In 2010, Verlinde elegantly demonstrated \cite{Verlinde2010On-the-Origin} that one can derive Newton's law of gravitation and Newton's second law relating acceleration and inertia by simple thermal considerations near a holographic screen surrounding a given region of spacetime. A similar result was derived by Padmanabhan \cite{Padmanabhan2010Surface} a month of two prior to Verlinde's paper.

B. Roy Frieden has been promulgating the Fisher-information based paradigm for physics (and indeed for all science) since at least the early 1990s \cite{Frieden1990Fisher,Frieden1995Erratum:,Frieden1995Lagrangians,Frieden1999Fisher-based,Frieden2002Black,Frieden2009Extreme}. In any discussion of the Universe as a computational system obeying information-theoretic laws, Frieden's work deserves a place of honor. An \href{http://www.fqxi.org/community/forum/topic/1816}{essay} by Rovelli submitted to the present competition also advocates an information based approach to physical law.

\section{Universal Quantum Gates}\label{sec:gates}

Classical computation is accomplished by means of Boolean gates such as AND, OR, XOR, NOT and NAND. These gates take one or more cbits (\emph{cbit}: ``classical bit'') $c \in \{0,1\}$ as input and the result of operating on the input results in output consisting of one or more cbits. In the quantum realm, our fundamental entities are quantum states $ \fullket{\Psi} \in \mc{H}$, in some Hilbert space, which serve as input and output. The computational gates take the form of unitary operators acting on $\mc{H}$. \emph{Programs} then consist of a number of these gates wired together in the form of a circuit, with the output of one set of gates serving as input for the next set.

An example illustrating the quantum circuit which takes three qubits $\{\fullket{0},\fullket{0},\fullket{0}\}$ as input and outputs an entangled state:
\begin{equation}\label{eqn:ghz-state}
  \fullket{\Psi_{GHZ}} = \frac{1}{\sqrt{2}}\left( \fullket{000} + \fullket{111} \right)
\end{equation}
known as the Greenberger-Horne-Zeilinger (GHZ) state. The circuit\footnote{example is taken from \href{http://www.texample.net/tikz/examples/quantum-circuit/}{texample.net}} which performs this operation is shown in the following figure:
\begin{figure}[h]
  \centerline{
    \begin{tikzpicture}[thick]
    %
    \tikzstyle{operator} = [draw,fill=white,minimum size=1.5em] 
    \tikzstyle{phase} = [fill,shape=circle,minimum size=5pt,inner sep=0pt]
    \tikzstyle{surround} = [fill=blue!10,thick,draw=black,rounded corners=2mm]
    %
    \node at (0,0) (q1) {\fullket{0}};
    \node at (0,-1) (q2) {\fullket{0}};
    \node at (0,-2) (q3) {\fullket{0}};
    %
    \node[operator] (op11) at (1,0) {H} edge [-] (q1);
    \node[operator] (op21) at (1,-1) {H} edge [-] (q2);
    \node[operator] (op31) at (1,-2) {H} edge [-] (q3);
    %
    \node[phase] (phase11) at (2,0) {} edge [-] (op11);
    \node[phase] (phase12) at (2,-1) {} edge [-] (op21);
    \draw[-] (phase11) -- (phase12);
    %
    \node[phase] (phase21) at (3,0) {} edge [-] (phase11);
    \node[phase] (phase23) at (3,-2) {} edge [-] (op31);
    \draw[-] (phase21) -- (phase23);
    %
    \node[operator] (op24) at (4,-1) {H} edge [-] (phase12);
    \node[operator] (op34) at (4,-2) {H} edge [-] (phase23);
    %
    \node (end1) at (5,0) {} edge [-] (phase21);
    \node (end2) at (5,-1) {} edge [-] (op24);
    \node (end3) at (5,-2) {} edge [-] (op34);
    %
    \draw[decorate,decoration={brace},thick] (5,0.2) to
	node[midway,right] (bracket) {$\frac{\fullket{000}+\fullket{111}}{\sqrt{2}}$}
	(5,-2.2);
    %
    %
    \end{tikzpicture}
  }
  \caption{
    A quantum circuit for producing a GHZ state using
    Hadamard gates and controlled phase gates.
  }
\end{figure}
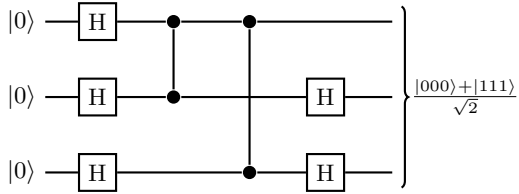
Here $H$ is the Hadamard gate which acts on a single qubit:
\begin{equation}
 H = \frac{1}{\sqrt{2}}\left( \begin{array}{ll}
				  1 & 1\\
				  1 & -1
			       \end{array}
			\right)
\end{equation}

In computation, both classical and quantum, there is a notion of a set of \emph{universal} gates which can be used to construct any computational circuit. For classical computation that is accomplished by the NAND gate in terms of which all the others gates (AND, OR, XOR, AND, NOR) can be constructed.

In the quantum arena, one choice of a universal gate is the CNOT gate which acts on two input qubits. The Hilbert space $H_1$ associated with one qubit is two-dimensional with basis $\{\fullket{0},\fullket{1}\}$. The two-qubit Hilbert space is the tensor product space $H_2 = H_1 \otimes H_1 $, spanned by the basis vectors $\{ \fullket{00},\fullket{01},\fullket{10},\fullket{11}\}$. Operators on $H_2$ are therefore represented by $4\times 4 $ matrices. The CNOT gate acting on elements of $H_2$ is given by \cite{Kauffman2004Braiding}:
\begin{equation}\label{eqn:cnot-gate}
  CNOT =  \left(
	  \begin{array}{cccc}
	   1 & 0 & 0 & 0 \\
	   0 & 1 & 0 & 0 \\
	   0 & 0 & 0 & 1 \\
	   0 & 0 & 1 & 0
	  \end{array}
	  \right)
\end{equation} 
Any quantum computational circuit can be constructed solely in terms of the two-qubit CNOT gate assisted by various single qubit \emph{phase} gates.

\section{Quantum Computation with Braids}\label{sec:topo-comp}

The field of topological quantum computation is based on the recognition that one can exploit the robustness to external perturbations of topological degrees of freedom in $2+1$ dimensional systems, such as the quantum hall fluid, in order to perform fault-tolerant quantum computation \cite{Freedman2000A-modular,Freedman2000Simulation,Freedman2002Topological,Bonesteel2005Braid,Nayak2008NonAbelian}.

\begin{figure}[h]
      \begin{center}
      		{\includegraphics[height=25mm]{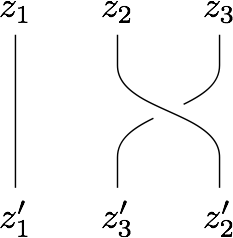}}
      		\caption{Braiding between two quasiparticles}
			\label{fig:braiding1}
	\end{center}
\end{figure}

For instance, in Fig. \ref{fig:braiding1}, $z_1, z_2$ and $z_3$ are three quasiparticles embedded in a 2DEG. Time evolution is represented by the downwards vertical axis. The state of such a system with $n$ such quasiparticles at a given instant of time $t$ can be written as:
\begin{equation}
	\fullket{\Psi(t)} = \fullket{z_1,\ldots,z_i,\ldots,z_j,\ldots,z_n}
\end{equation}
where $z_i$ are the two-dimensional co-ordinates of each quasiparticle. Such a state is insensitive to changes of the co-ordinates $z_i \rightarrow z_i + \delta z_i$, which do not interchange the positions of pairs of quasiparticles. However, moving any two such quasiparticles ($z_2$ and $z_3$ in the illustration below \ref{fig:braiding1}) around each other - and thus interchanging their positions - causes the state to pick up a phase factor:
\begin{align}
	\fullket{\Psi(t')} & = \fullket{z_1,\ldots,z_j,\ldots,z_i,\ldots,z_n} \nonumber \\ 											& = e^{-i \theta_{ij}} \fullket{z_1,\ldots,z_i,\ldots,z_j,\ldots,z_n}
\end{align}
The net effect of such an operation is to perform a unitary transformation $\mb{R}$ on the Hilbert space of the quasiparticle pair $\{z_i,z_j\}$. As we will show in the next section, one can construct a set of such braiding moves, which are equivalent to acting upon the pair of qubits - with each quasiparticle encoding a single qubit - with a CNOT gate \ref{eqn:cnot-gate}. Then, as mentioned above, in combination with certain single-qubit phase gates, the resulting system provides a complete setup for performing universal quantum computation.

In order to compute the expectation value of various braiding operations, such as the one depicted in Fig. \ref{fig:braiding1}, we connect the top and bottom ends of the strands as shown in the following figure:

\begin{figure}[h]
      \begin{center}
      		{\includegraphics[height=32mm]{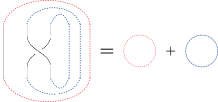}}
      		\caption{Closing the braid to form a collection of links and knots}
			\label{fig:closed_braid}
	\end{center}
\end{figure}

We are then left with a collection of links and knots. In the above simple case, the braid reduces to two disconnected zero-knots (or simply ``circles''). The quantity we are then interested in calculating is the partition function of the topological field theory representing the physical system at hand, over the given knot configuration. The prototypical topological qft (tqft) is (abelian or non-abelian as the case may be) Chern-Simons theory whose action is given by:
\begin{equation}\label{eqn:cs-action}
	S_{cs} = \int d^2 x\,dt \,\, A \wedge \mb{d} A + \frac{2}{3} A \wedge A \wedge A
\end{equation}
where $A$ is a gauge field which encodes the dynamics of the quasiparticles. Let $\mc{K}$ denote the resulting collections of links and knots. Locally ${\cal K}$ is just a one-dimensional path. Given a \emph{lie-algebra valued} gauge connection $ A_{a}^i $ on our $2+1$ dimensional manifold $ {\cal M} $, one can evaluate the holonomy of $ A_{a}^i $ along ${\cal K}$:
\begin{equation}\label{eqn:knot-holonomy}
	H_{\cal K}  = Tr \left[ \mc{P} \exp \left( -i \int d\gamma A_{a}^i t_i x^a(\gamma)  \right) \right]
\end{equation}
where $ \mc{{P}} $ denotes that the integral is ``path ordered'', $\gamma$ parametrizes points along the curve, $x^a(\gamma)$ is the tangent vector to the curve at a given point, and $t_i$ are the generators of the relevant lie-algebra. The trace is taken over the lie-algebra indices. $H_{\mc{K}}$ is the relevant gauge invariant observable and its expectation value $\expect{H_{\mc{K}}}$ for the knot(s) formed by closing the braid can be written in terms of the partition function for Chern-Simons theory:
\begin{equation}
	W_{\cal K} = \expect{H_{\cal K}} = \int D \mc{A} \; \exp \left(-i \beta S_{cs} \right) H_{\cal K}
\end{equation}
Now, as shown by Witten in his ground-breaking work \cite{Witten1989Quantum}, the above quantity is equivalent to the value of the polynomial invariant known as the Jones polynomial evaluated on the given knot(s). It is these expectation values, that will ultimately correspond to the physical results of the quantum computation expressed in terms of a series of braiding operations on the set of quasiparticles in the two-dimensional plane.

\textbf{Quantization of Physical Observables}: Physical observables should be invariant under ``small'' diffeomorphisms - \emph{i.e.}, those which do not change the homotopy class of the set of punctures on the plane. Quantities such as $W_\mc{K}$ do not change under small diffeomorphisms. Since the space of homotopy classes is a discrete set, expectation values of gauge invariant observables will also take values in a discrete set and should therefore be quantized.


\section{Yang-Baxter Equation}\label{sec:yang-baxter}


So far we have only spoken of the braiding operation, and its effect on the quasiparticle wavefunction, in the abstract. Let us now determine concretely the possible forms of the unitary two-qubit operation $\mb{R}$. But, first, let us describe the braid group in more detail.

The braid group on $n$ strands, is a discrete, infinite group with $n$ generators $\{\sigma_i,\ldots,\sigma_n\}$, where the effect of $\sigma_i$ is to braid the $i$\supersc{th} and $(i+1)$\supersc{th} strands, such that the $i$\supersc{th} strand passes \emph{over} the $(i+1)$\supersc{th} strand\footnote{Note that whether a strand passes ``over'' or ``under'' another depends on the particular projection of the three dimensional braid onto a two-dimensional surface, but once a projection is fixed, the sense of ``over'' and ``under'' is the same for all strands.}. $\sigma_i^{-1}$ is the inverse of this generator causing the two strands to braid in the opposite sense. Any two generators $\sigma_i$ and $\sigma_j$, commute as long as $|i-j| > 1$. Neighboring generators satisfy the relation:
\begin{equation}\label{eqn:braid-algebra}
	\sigma_i \sigma_{i+1} \sigma_i = \sigma_{i+1} \sigma_i \sigma_{i+1}
\end{equation}
This can be schematically represented as the following relation:
\begin{equation}\label{eqn:yang-baxter-braid}
	\vcenter{\hbox{\begin{tikzpicture}
\braid [
		number of strands=3,  
		line width=1pt,
		yscale=1
]
	(braid_1) a_1 a_2 a_1;
\end{tikzpicture}}} = \vcenter{\hbox{\begin{tikzpicture}
\braid [
		number of strands=3,  
		line width=1pt,
		yscale=1
]
	(braid_1) a_2 a_1 a_2;
\end{tikzpicture}}}
\end{equation}
between the $i$\supersc{th}, $(i+1)$\supersc{th} and $(i+2)$\supersc{nd} strands (labeled from left to right, with the other strands being suppressed in the illustration). This relationship is called the \emph{braided} Yang-Baxter equation. We note that the action of $\sigma_i$ on the $i$\supersc{th} and $(i+1)$\supersc{th} strands corresponds to acting with the previously mentioned unitary operation $\mb{R}$ on the pair of qubits represented by the two strands. This statement can be formalized as follows.

At the ends of each strand, a copy of a vector space $V$ is attached. A braid with $n$ strands represents the $n$-fold tensor product $\prod_{i=1}^{n} V$ of $n$ copies of $V$. For the Yang-Baxter equation we have three strands, i.e. we are working with $V \otimes V \otimes V$. The operator $\mb{R}$ acts on $V \otimes V$ and can be represented as a braiding operation between two strands. For example, let $V$ be a vector-space, such as the Hilbert space of a spin-$\onehalf$ system. Let $\mb{R}$ be an automorphism on $V \otimes V$:
\begin{equation}
  \mb{R} : V \otimes V \rightarrow V \otimes V
\end{equation}
Let $\mathbf{1}$ denote the identity operator on $V$.
Then $\mb{R}$ is said to satisfy the \emph{algebraic} Yang-Baxter equation if the following holds:
\begin{equation}\label{eqn:yang-baxter-algebraic}
  (\mb{R} \otimes \mathbf{1})(\mathbf{1} \otimes \mb{R})(\mb{R} \otimes \mathbf{1}) = (\mathbf{1} \otimes \mb{R}) (\mb{R} \otimes \mathbf{1})(\mathbf{1} \otimes \mb{R})
\end{equation}
Each term in parenthesis in the above equation is an operator acting on the tensor product space $V \otimes V \otimes V$. This relationship can be represented schematically as shown in \ref{eqn:yang-baxter-graphic}:  
\begin{equation}\label{eqn:yang-baxter-graphic}
  \vcenter{\hbox{\includegraphics[height=50mm]{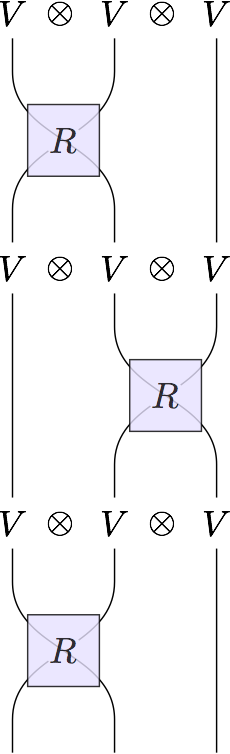}}} \quad = \quad \vcenter{\hbox{\includegraphics[height=50mm]{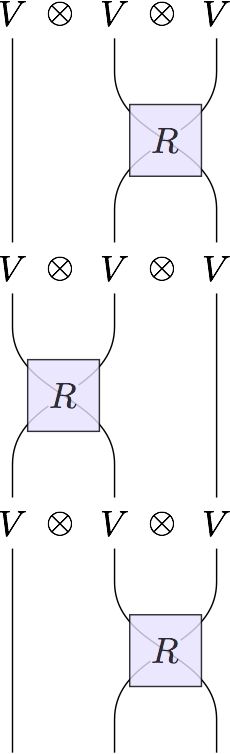}}}
\end{equation}
Thus in on the \emph{l.h.s.} of \ref{eqn:yang-baxter-graphic}, the top-most segment represents the action of $(R\otimes \mathbf{1})$ on $V \otimes V \otimes V$. The fact that $\mb{R}$ acts only on the first two copies of $V$ is represented by the braiding between the respective strands, and the identity operation on the third copy is simply represented by a straight line. Similarly the remaining two segments on the \emph{l.h.s.} show the action of $(\mathbf{1}\otimes R)$, followed once again by $(R\otimes \mathbf{1})$. In this manner the \emph{l.h.s.} of \ref{eqn:yang-baxter-graphic} is a graphical representation of the \emph{l.h.s.} of \ref{eqn:yang-baxter-algebraic}. Solutions of \ref{eqn:yang-baxter-algebraic} will yield the allowed forms of the unitary operation $\mb{R}$.

The crux of this essay lies in the fact that as shown by Kauffman and Lomonaco in \cite{Kauffman2004Braiding}, unitary solutions of \ref{eqn:yang-baxter-algebraic} corresponds to the CNOT operator \ref{eqn:cnot-gate}, mentioned in Sec. \ref{sec:gates}.

\section{Bilson-Thompson Model}\label{sec:bilson}

In 2005, Sundance Bilson-Thompson put forward an innovative approach towards understanding the structure of elementary particles in terms of the so-called \emph{preon model}, which utilized three-strand braids to construct the first generation of leptons of the standard model.
\begin{figure}[h]
      \begin{center}
      		{\includegraphics[height=37mm]{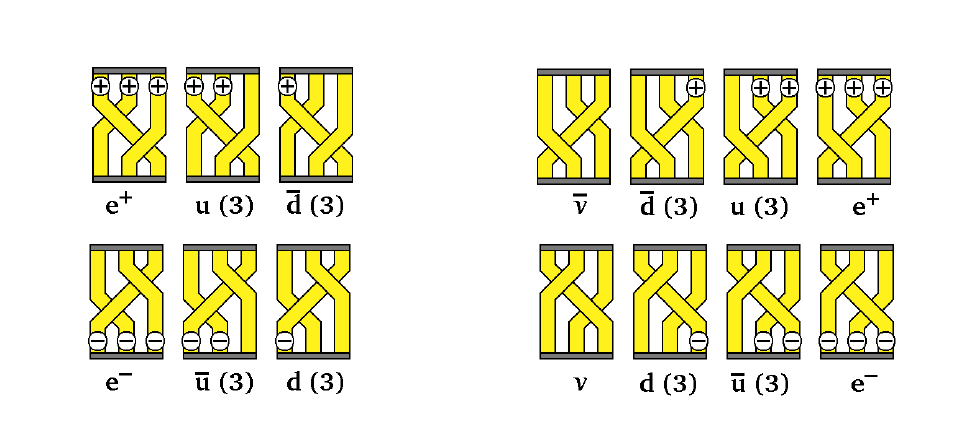}}
      		\caption{Bilson-Thompson's proposed model of elementary particles. Figure credit \cite{Bilson-Thompson2005A-topological}}
      \label{fig:bt_model}
	\end{center}
\end{figure}
Whereas, the braids we have considered so far have strands which are one-dimensional lines, the preon model utilizes so-called \emph{framed} ribbons instead. Allowing the strands to have a width, implies that in addition to the braiding in-between different strands, one can also have single-strand operations which correspond to putting a $180\deg$ ``twist'' in the strand. Such twists can have two senses depending on whether we twist the strand ``clockwise'' or ``anti-clockwise''. Bilson-Thompson identified such twists with adding a charge of $\pm 1/3$ on a strand (with the positive sign for clockwise twists and the negative for anti-clockwise twists).

The three-strand braid group $B_3$ has only one non-trivial element given by $\sigma_1 \sigma_2^{-1}$ in terms of the braid generators. With the additional degree of freedom provided by the twists, one can construct a map between all elements of the \emph{framed} braid group $\mf{B}_3$ and the first generation of leptons as shown in Fig. \ref{fig:bt_model}. $\mf{B}_n$ has the same set of generators $\{\sigma_i\}$ as $B_n$, augmented by the single-strand operators $\{h_i^{\pm}\}$ which generate twists on the framed ribbons. As mentioned at the end of Sec. \ref{sec:gates}, in order to perform universal quantum computation, we need only one operation - the CNOT - augmented by certain single-qubit phase gates. The braid group with one-dimensional strands $B_n$, naturally has the former ingredient, arising from the solutions of the algebraic Yang-Baxter equation. Adding framing to the strands can then supply us with the needed set of single-qubit phase gates and we suggest (without proof) that this is precisely the role played by the twist operators $h_i^\pm$.

\section{Elementary particles as information processors}\label{sec:processors}

There is as yet, no concrete theoretical (let alone experimental) evidence that Bilson-Thompson's braid model is an accurate representation of reality; though, by now, a substantial body of work has been put forward \cite{Bilson-Thompson2006Quantum,Wan2007Braid,Bilson-Thompson2008Particle,Wan2009Effective,Bilson-Thompson2009Particle,Vaid2010Embedding,Hackett2011aInvariants,Hackett2011bInvariants,Bilson-Thompson2012Emergent} in support of such a conjectured correspondence between elementary particles and the three-strand braid group $B_n$ or its framed counterpart $\mf{B}_3$.

However, the connections between topology, quantum computation and quantum gravity that emerge from following these ideas to their natural conclusion are far too compelling to be mere coincidence. The fact that Chern-Simons theory describes the behaviour of boundary surfaces in LQG is more than just a technical observation. In combination with the role played by Chern-Simons theory in topological quantum computation, we are lead to suggest an intimate relation between computation and the dynamics of LQG. The suggestion that spin-networks are inherently computational objects, is itself, not novel having been previously explored in \cite{Zizzi2001Quantum,Marzuoli2004Spin,Garnerone2006Quantum,Girelli2005Reconstructing,Livine2006Reconstructing,Kadar2008Braiding}.

What is novel in the present work is a concrete proposal for viewing the particles of the standard model as information processing objects, or more precisely, as gates for universal quantum computation. If the Universe is to be thought of as a quantum computer, then what could be more elegant than to suppose that spacetime provides the physical substratum for computation, on which reside elementary particles and gauge fields, with the former playing the role of the building blocks (\emph{hardware}) for computation, and the interaction rules for quantum fields playing the role of \emph{software} which determines how the computation is carried out.


\begin{acknowledgments}
I would like to thank my parents for unrelenting support in the face of professional and personal obstacles. A special thanks is owed to Sundance Bilson-Thompson for his support and encouragement over the years. Part of this work was done during a visit to the Perimeter Institute in Fall 2009. The core ideas in this essay were first presented at seminars at the Harish-Chandra Research Institute (HRI), Allahabad and the Indian Institute of Science (IISc), Bangalore in April, 2010.
\end{acknowledgments}

\bibliographystyle{JHEP3}

\bibliography{compuni.bib}

\appendix

\section{``Large'' Gauge Transformations and Braided Spin Networks}\label{sec:large_gauge}

In any gauge theory there are two kinds of gauge transformations, invariance under which determines the physical state space of the theory. These are the \emph{small} (or \emph{connected}) and \emph{large} (or \emph{dis-connected}) gauge transformations. A small gauge transformation is one that can be continuously deformed into the identity, whereas large gauge transformations cannot. Small gauge transformations lie in the component of the gauge group which contains the identity and large transformations lie in the disconnected components of the gauge group.

An example is given by the Wu-Yang monopole solution in the Yang-Mills theory of the gauge group $SU(2)$ on $\mbb{R}^3$. We specify the state of the field by specifying at each point on a spacetime manifold an element of $\mf{su}(2)$ - \emph{i.e.} an $SU(2)$ ``spin''. The spin-vector at a given location is related to its neighbors by an element of $SU(2)$. Now, recall that $\mbb{R}^3$ can be compactified and mapped to a three-dimensional sphere $S^3$ by adding the point at infinity. Likewise, the manifold of $SU(2)$ is also homeomorphic to a 3-sphere. Gauge transformations are $SU(2)$ valued functions on the spacetime manifold, i.e. they are maps $\phi:S^3 \rightarrow S^3$. These maps lie in disjoint homotopy-inequivalent classes. Members of one equivalence class cannot be smoothly deformed into members of another equivalence class without introducing singularities. \emph{Large} gauge transformations are those which connect maps in different homotopy classes.

Likewise, the \emph{full} symmetry group of a $3+1$ dimensional manifold consists not only of small diffeomorphisms - \emph{i.e.}, infinitesimal translations, rotations and scale transformations - but also by large gauge transformations. In fact, it is known \cite{Friedman1982Half-integral} that the existence of large diffeomorphisms of three-manifolds, \emph{i.e.} those which are not asymptotically trivial and hence cannot be continuously deformed into the identity, implies the existence of states in the physical Hilbert space labeled by half-integer angular momenta. In other words, half-integral spin arises in quantum gravity purely from considerations of invariance of the state space under large diffeomorphisms. From these considerations it would appear that large diffeomorphisms should also play an important role in determining the physical state-space of loop quantum gravity.

\begin{figure}[h]
      \begin{center}
      {\includegraphics[height=40mm]{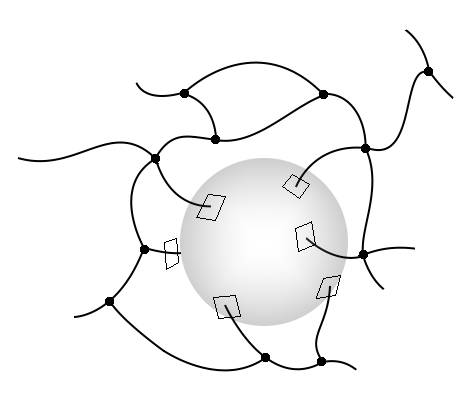}}
      \caption{The quantum state of the surface of a black hole in LQG is specified by a spin-network graph and the way its edges puncture the boundary of the black hole.}
      \label{fig:black-hole-area}
\end{center}
\end{figure}

We can illustrate the effect of small and large gauge transformations in the context of LQG, where states of quantum geometry are given by spin-networks which are graphs whose edges are labeled with representations of $SU(2)$ and whose vertices are labeled by \emph{intertwiners} invariant tensors in the tensor Hilbert space of all the spins coming into a given vertex. These graphs are embedded in a background manifold which only has the structure of a topological space. Two-dimensional surfaces in this manifold are endowed by area at the locations where a surface is pierced by an edge. This allows us to construct a quantum geometric description of the surface of a black hole as shown in Figure \ref{fig:black-hole-area}.

\begin{figure}[h]
      \begin{center}
      {\includegraphics[height=30mm]{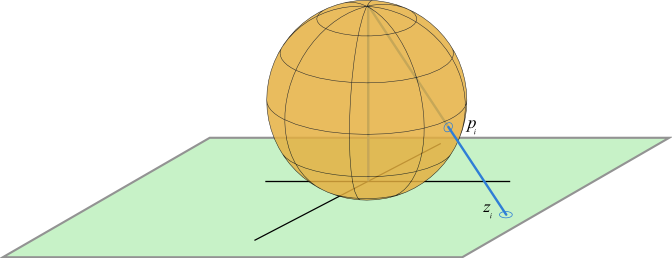}}
      \caption{Stereographic projection of the 2-sphere onto the complex plane}
      \label{fig:stereographic}
\end{center}
\end{figure}

\begin{figure}[h]
      \begin{center}
      {\includegraphics[height=40mm]{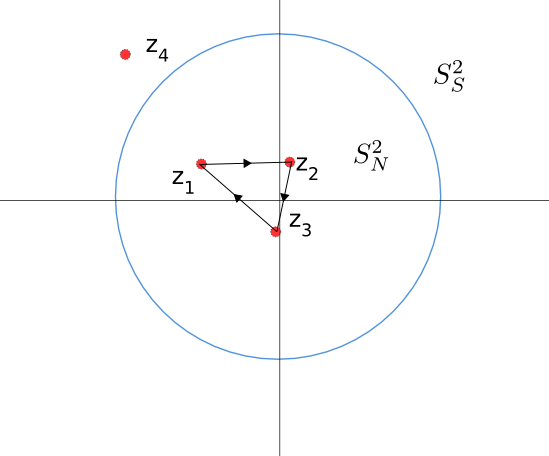}}
      \caption{Projection of the four-punctured sphere onto the complex plane. $S^2_N$ and $S^2_S$ are the northern and southern hemispheres of the sphere. One can always arrange to have three of the punctures lie in one hemisphere by means of a $\sltwoc$ transformation \cite{Penrose1988Spinors}.}
      \label{fig:four-punctures}
\end{center}
\end{figure}

Let us consider the simplest case\footnote{The cases where a 2-sphere has only one, two or three punctures do not uniquely specify a state of quantum geometry. The four-valent case is the first non-trivial example.} consisting of a surface punctured in four places by a spin-network. By means of a stereographic projection one can map the surface of a 2-sphere to the complex plane as shown in Figure \ref{fig:stereographic}. A 2-sphere with four-punctures can then be represented as shown in Figure \ref{fig:four-punctures}.

\begin{figure}[h]
	\begin{center}
	\subfloat{
      		\includegraphics[height=30mm]{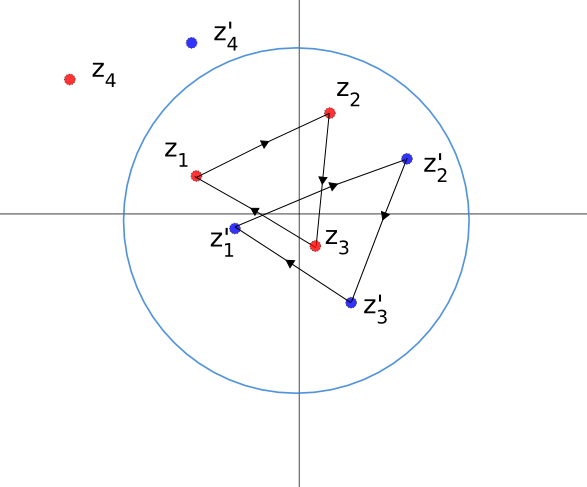}
	}
	\subfloat{
		\includegraphics[height=30mm]{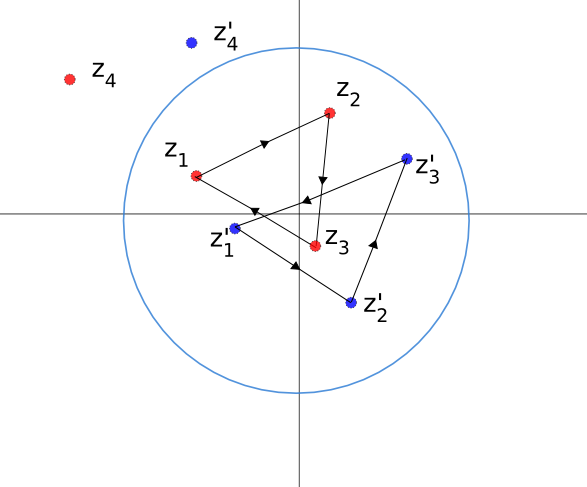}
	}
	\caption{``Small'' diffeomorphisms move the punctures around but do not change the relative orientation (\emph{left}). ``Large'' diffeomorphisms have the effect of exchanging the locations of two punctures (\emph{right}), in this case those labeled $z_2$ and $z_3$.}
    \label{fig:large-small-diffeos}
    \end{center}
\end{figure}

The effect of small and large diffeomorphisms, acting in the bulk, on the surface states of the four-valent sphere is shown in Figure \ref{fig:large-small-diffeos}. The requirement of invariance of physical states under small diffeomorphisms implies that it is only the values of the spins on the edges which determine the state and not the location of the punctures, since the co-ordinates on the sphere do not have any physical meaning. However, one is still left with the redundancy induced by large diffeomorphisms which cause two punctures to be exchanged. Rotating one puncture around other has the effect of braiding their respective edges around each other.

\begin{figure}[h]
      \begin{center}
      {\includegraphics[height=25mm,angle=0]{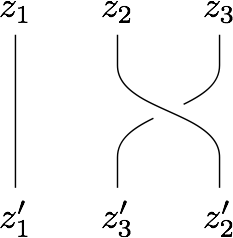}}
      \caption{Large diffeomorphisms lead to braiding of spin-network edges around each other. (image credit: the latex ``braids'' package by Andrew Stacey)}
      \label{fig:braiding}
\end{center}
\end{figure}

This is illustrated schematically in Fig. \ref{fig:braiding}.  In order to fully incorporate the requirement of diffeomorphism invariance in quantum gravity it is not sufficient to only the usual spin-networks. Though spin-networks do satisfy invariance under ``small'' diffeomorphisms, they neglect to take into account the action of large diffeomorphisms who effect is to braid edges of the spin-network. Taking this requirement fully into account leads us to the Yang-Baxter equation which allows us to trace a correspondence between the preon model and quantum computation.

\end{document}